\documentclass[aps, prb, floatfix, showpacs, preprint, preprintnumbers, superscriptaddress, amsmath, amssymb]{revtex4-1}

\usepackage{graphicx}% Include figure files
\usepackage{times}
\usepackage{dcolumn}
\usepackage{amsfonts}
\usepackage{bm}
\begin{document}
%\begin{CJK*}{GBK}{kai}

\preprint{v.8}

\title{\boldmath Unusual slow magnetic fluctuations and critical slowing down in Sr$_{2}$Ir$_{1-x}$Rh$_{x}$O$_{4}$}

\author{C. Tan}
\author{Z. F. Ding}
\author{J. Zhang}
\author{Z. H. Zhu}
\affiliation{State Key Laboratory of Surface Physics, Department of Physics, Fudan University, Shanghai 200433, China}
\author{O. O. Bernal}
\affiliation{Department of Physics and Astronomy, California State University, Los Angeles, California 90032, USA}
\author{P. C. Ho}
\affiliation{Department of Physics, California State University, Fresno, California 93740, USA}
\author{A. D. Hillier}
\affiliation{ISIS Facility, STFC Rutherford Appleton Laboratory, Harwell Science and Innovation Campus, Chilton, Didcot, Oxon, United Kingdom}
\author{A. Koda}
\affiliation{Institute of Materials Structure Science, High Energy Accelerator Research Organization(KEK), Tokai, Ibaraki, 319-1106, Japan.}
\author{H. Luetkens}
\affiliation{Laboratory for Muon-Spin Spectroscopy, Paul Scherrer Institut, 5232 Villigen, Switzerland}
\author{G. D. Morris}
\affiliation{Centre for Molecular and Materials Science, TRIUMF, Vancouver, British Columbia V6T 2A3, Canada}
\author{D. E. MacLaughlin}
\affiliation{Department of Physics and Astronomy, University of California, Riverside, California 92521, USA}
\author{L. Shu}
\thanks{Corresponding author: leishu@fudan.edu.cn}
\affiliation{State Key Laboratory of Surface Physics, Department of Physics, Fudan University, Shanghai 200433, China}
\affiliation{Collaborative Innovation Center of Advanced Microstructures, Nanjing 210093, China}
\date{\today}
\begin{abstract}
Hidden magnetic order in the correlated iridate~Sr$_2$Ir$_{1-x}$Rh$_x$O$_4$, $x = 0.05$ and 0.1, has been studied using muon spin relaxation spectroscopy. In zero field (ZF) and weak longitudinal fields (LF) ($\lesssim 2$~mT), the muon spin relaxation data indicate that static and dynamic local fields coexist at each muon site, and can be well described by exponentially-damped static Lorentzian Kubo-Toyabe functions. The ZF relaxation rate is dominated by the static field distribution, and a broad relaxation rate maxima at 175~K for $x = 0.1$ in ZF is attributed to muon diffusion and trapping. For $\text{LF} \gtrsim 2$~mT the static rate is completely decoupled, and the exponential decay is due to dynamic spin fluctuations. The temperature dependences of the relaxation rates exhibit maxima at 215~K ($x = 0.05$) and 175~K ($x = 0.1$), in agreement with previous second-harmonic generation and polarized neutron diffraction determinations of transition temperatures to a hidden-order state. The maxima are most likely due to critical slowing down of electronic spin fluctuations. The field dependences of the dynamic spin fluctuation rates can be well described by the Redfield relation, from which the rms width~$B_\mathrm{loc}^\mathrm{rms}$ and correlation time~$\tau_\mathrm{c}$ of the fluctuating field are obtained. Values of $\tau_c$ are in the range of~1.5--4~ns for $x = 0.1$ and shorter than 2~ns for $x = 0.05$, suggesting an increase with increasing Rh concentration. Values of $B_\mathrm{loc}^\mathrm{rms}$ are on the order of 1~mT, consistent with the polarized neutron diffraction cross-section.

\end{abstract}

%\pacs{75.10.Hk, 05.10.Ln, 64.60.Cn, 11.15.Ha }% PACS, the Physics and Astronomy
 %Classification Scheme.

\maketitle

\section{INTRODUCTION}

Iridium oxides have attracted growing attention because their physical properties can be significantly influenced by the combination of strong spin-orbit interactions and electron correlations. Sr$_2$IrO$_4$, an archetypal $J_\mathrm{eff}=1/2$ Mott insulator, has structural, electronic, and magnetic properties similar to those of the $S=1/2$ Mott insulator La$_{2}$CuO$_{4}$~\cite{Wang2011Twisted, Watanabe2013Monte, Meng2014Odd, Ye2013Magnetic, Dhital2013Neutron}, with an antiferromagnetic (AFM) transition at N\'eel temperature~$T_N = 230$~K~\cite{Cao1998Weak, Kim2008Novel, Franke2011Measurement}. Fermi arcs and a V-shaped low-energy gap in electron-doped Sr$_2$IrO$_4$ remarkably resemble the properties of hole-doped cuprate superconductors~\cite{Kim2014Fermi, Torre2015Collapse, Yan2015Electron, Kim2015Observation}. Theoretical predictions and empirical experimental observations suggest that doped Sr$_2$IrO$_4$ is a promising system for comparison with cuprates, and may even possess new states of matter~\cite{Wang2011Twisted,Chikara2017Charge}.

Recently an odd-parity hidden order that breaks spatial inversion and rotational symmetry was inferred from an optical second harmonic generation (SHG) study of undoped and Rh-doped Sr$_2$IrO$_4$~\cite{Zhao2015Evidence} above $T_N$\@. The emergence of hidden magnetic order reduces the C$_4$ rotational symmetry of rotational anisotropy SHG pattern to C$_1$. This has two-fold rotational symmetry, which is significantly different from that of antiferromagnetic (AFM) order. Polarized neutron diffraction (PND) experiments~\cite{Jeong2017Time-reversal} further revealed that this order breaks time-reversal symmetry, while preserving translation symmetry of the lattice. Magnetic diffraction was observed at $(1,1,2)$ in momentum space whereas AFM order reflections would appear at $(1,0,L{=}1, 2)$. The broken symmetries of this hidden magnetic order are consistent with the loop-current model proposed~\cite{Varma1997NonFermi, *Varma2006Theory} to explain exotic properties of the pseudogap state in cuprates~\cite{Timusk1999pseudogap}. Sr$_2$Ir$_{1-x}$Rh$_x$O$_4$ differs from the cuprates in that it exhibits an AFM phase, but its behavior above $T_N$ is similar to that of the cuprate pseudogap phases.

The nature of the pseudogap state is a fascinating puzzle in condensed matter physics. The observation of spontaneous breaking of time-reversal symmetry in Bi$_2$Sr$_2$CaCu$_2$O$_{8+\delta}$~\cite{Kaminski2002Spontaneous} provided evidence that the pseudogap state is a novel phase rather than a crossover. The pioneering PND studies of Fauqu\'e \emph{et~al.}\ on YBa$_2$Cu$_3$O$_{6+x}$ (YBCO)~\cite{Fauque2006Magnetic} revealed intra-unit-cell (IUC) magnetic order (i.e., order that preserves lattice translational symmetry) in the pseudogap phase. Similar pseudogap IUC magnetic order was also found in HgBa$_2$CuO$_{4+\delta}$~\cite{Li2008Unusual} and Bi$_2$Sr$_2$CaCu$_2$O$_{8+\delta}$~\cite{De2012Evidence}.

However, the absence of static magnetic fields with the expected amplitude from NMR~\cite{Mounce2013Absence, Wu2015Incipient} and muon spin relaxation/rotation ($\mu$SR)~\cite{MacDougall2008Absence, Sonier2009Detection, Huang2012Precision, Pal2016Investigation} experiments raised doubts concerning the existence of IUC magnetic order. Recent $\mu$SR experiments of Zhang \emph{et~al.}~\cite{Zhang2018Discovery} revealed a slowly-fluctuating field in the pseudogap phase of YBCO with a correlation time $\tau_\mathrm{c}$ on the order of 10$^{-9}$~s. These fluctuations are therefore quasistatic on the neutron time scale ($10^{-11}$~s)~\cite{Jeong2017Time-reversal}. They explain the absence of a static field, and are consistent with a theoretical treatment involving domains of fluctuating IUC order~\cite{Varma14Pseudogap}.

The $\mu$SR technique~\cite{schenck1985muon, brewer1994encyclopedia, yaouanc2011muon} is unmatched in its sensitivity to magnetic field on the local (atomic) scale. $\mu$SR is capable of detecting static magnetic moments in the range of 0.001--0.01$\mu_\mathrm{B}$, and can measure correlation times in the range~$10^{-4}$--$10^{-12}$~s, depending on the magnitude of fluctuating fields at muon sites. Since the hidden-order phase in Sr$_2$Ir$_{1-x}$Rh$_x$O$_4$ resembles that in the pseudogap state of cuprates~\cite{Zhao2015Evidence, Jeong2017Time-reversal}, hidden-order magnetism might also fluctuate with a comparable time scale in Sr$_2$Ir$_{1-x}$Rh$_x$O$_4$. $\mu$SR should be a suitable probe to detect such fluctuations.

We report the observation of slow spin dynamics with a time scale of 10$^\textsc{-9}$~s, comparable to that in cuprates~\cite{Zhang2018Discovery}, using zero-field (ZF) and longitudinal-field (LF) $\mu$SR techniques. The results suggest a similar origin of electronic spin dynamics. The temperature dependences of ZF and LF relaxation rates exhibit maxima at temperatures~$T^\ast$ consistent with the hidden-order temperatures observed in SHG and PND experiments~\cite{Zhao2015Evidence,Jeong2017Time-reversal}. Values of correlation times~$\tau_c$ and rms fluctuating fields~$B_\mathrm{loc}^\mathrm{rms}$ are comparable to those in YBCO~\cite{Zhang2018Discovery}. We interpret these results as further evidence for the hidden order revealed by the SHG and PND studies.

\section{METHODS} \label{sec:methods}

\subsection{Samples, characterization, facilities} \label{samples}

\begin{figure} [ht]
 \includegraphics[width=0.45\textwidth]{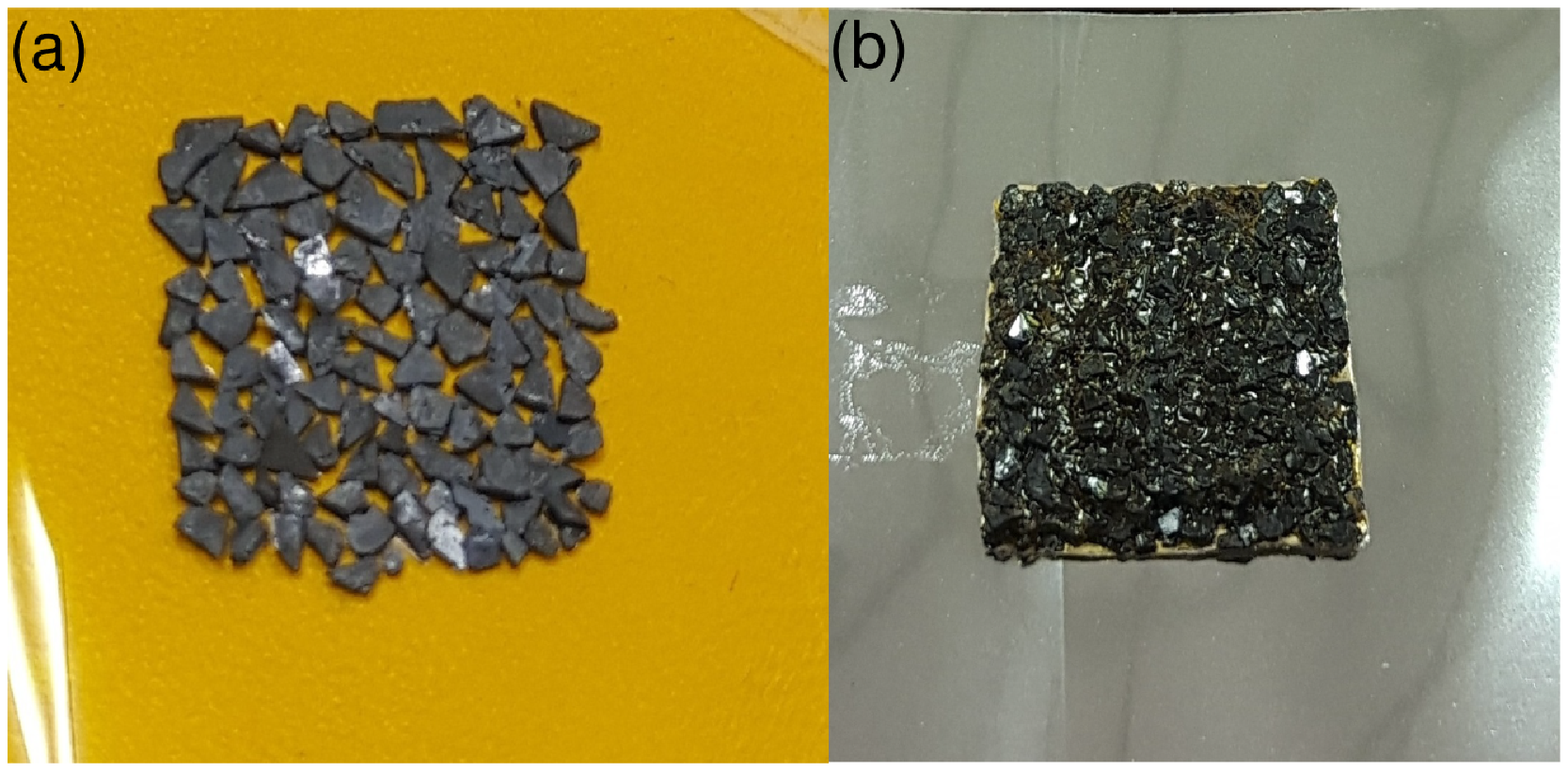}
 \caption{\label{fig:photo} (a) Sr$_2$Ir$_{0.95}$Rh$_{0.05}$O$_4$ and (b) Sr$_2$Ir$_{0.9}$Rh$_{0.1}$O$_4$ single-crystal mosaic samples for $\mu$SR experiments. Crystals of Sr$_2$Ir$_{0.95}$Rh$_{0.05}$O$_4$ were mounted on plastic tape with $ab$ planes in the layer, the overall size is 10$\times$10 mm$^2$. Crystals of Sr$_2$Ir$_{0.9}$Rh$_{0.1}$O$_4$ were mounted on a silver plate in three layers, with $ab$ planes in each layer. The overall size is 11$\times$11 mm$^2$.}
\end{figure}

Single crystals of Sr$_{2}$Ir$_{1-x}$Rh$_{x}$O$_{4}$, $x = 0.05$ and 0.1, were grown from off-stoichiometric quantities of SrCl$_2$, SrCO$_3$, IrO$_2$, and RhO$_2$ using a self-flux method. Details of the method are described elsewhere~\cite{Sung2016Crystal}. For $x = 0.05$ the average size of the crystallites is 1$\times$1$\times$0.2 mm$^3$, with the 1$\times$1 surface in the \emph{ab} plane~\cite{Qi2012Spin}; the crystallites are somewhat smaller for $x = 0.1$. For each sample a mosaic of single crystallites was mounted with $c$ axes oriented roughly normal to the muon beam direction and $ab$ axes unaligned. Figure~\ref{fig:photo} shows photos of the samples.

Powder x-ray diffraction, magnetic susceptibility, and resistivity measurements were performed on a Bruker DISCOVER 8 diffractometer, a Quantum Design SQUID magnetometer, and a Quantum Design physical properties measurement system, respectively. ZF-, LF-, and weak transverse field (wTF) $\mu$SR experiments were performed using the ISIS EMU spectrometer at Rutherford Appleton Laboratory, Didcot, U.K., the S1 spectrometer at J-PARC, Tokai, Japan, and the GPS spectrometer at the Paul Scherrer Institute, Villigen, Switzerland.

\subsection{The $\bm{\mu}$SR technique} \label{sec:musrtech}

In the time-differential $\mu$SR technique~\cite{schenck1985muon, brewer1994encyclopedia, yaouanc2011muon}, the time evolution of the ensemble muon spin polarization~$G(t)$ in a sample, where $t$ is the time after muon implantation, is obtained from the asymmetry time spectrum~$A(t)$ of beta-decay positron count rates. In general $A(t) = A_0G(t)$, where the initial asymmetry~$A_0$ depends on spectrometer details but is roughly 0.2. In the ZF- and LF-$\mu$SR techniques the positron counters are oriented to detect the asymmetry in the initial muon polarization direction, so that $G(0) = 1$.

The behavior of local fields at muon sites due to nuclear and electronic magnetism is reflected in the form of $G(t)$. Relaxation (loss of muon polarization) is due to dephasing of muon precession in a distribution of static local fields, dynamic fluctuations of local fields, or a combination of these. Often the form of the static local field distribution can be determined.

Nuclear magnetism usually results in dipolar fields that are quasistatic (slow compared to the relevant time scale)~\cite{KuTo67, Hayano19179Zero}. For randomly-oriented quasistatic Gaussian and Lorentzian field distributions, the static ZF muon spin polarization functions~$G(t)$ are given by~\cite{Hayano19179Zero}
\begin{widetext}
\begin{equation} \label{Eq:KTGauss}
 G_\mathrm{ZF}^\mathrm{Gauss}(\sigma_s,t) = \frac{1}{3}+\frac{2}{3}\left[1-(\sigma_s t)^2\right]\exp\left[-\textstyle{\frac{1}{2}}(\sigma_s t)^2\right] \quad (\mathrm{Gaussian})
\end{equation}
and~\cite{Uemura1985Muon}
\begin{equation} \label{Eq:KTLor}
 G_\mathrm{ZF}^\mathrm{Lor}(\lambda_s,t) = \frac{1}{3}+\frac{2}{3}(1-\lambda_s t)\times \exp(-\lambda_s t) \quad (\mathrm{Lorentzian}),
\end{equation}
\end{widetext}
respectively. Here $\sigma_s/\gamma_\mu$ and $\lambda_s/\gamma_\mu$ are the rms width and HWHM, respectively, of the field distributions.

These functions are examples of so-called Kubo-Toyabe (KT) relaxation~\cite{KuTo67}. Static muon spin relaxation can be ``decoupled'' from the static fields by applying a sufficiently strong longitudinal field, since then the resultant field is nearly parallel to the initial muon polarization and there is little muon precession. Observation of the late-time recovery of these functions is limited by the muon lifetime to cases where the relaxation is sufficiently fast, and the recovery is to 1/3 only for randomly oriented local fields.

Dynamic muon spin relaxation is caused by fluctuating local fields~$B_\mathrm{loc}(t)$ at muon sites. $G(t)$ depends on properties of $B_\mathrm{loc}(t)$, in particular the correlation time~$\tau_{c}$ that characterizes its stochastic time dependence. The relaxation is in the so-called ``motional narrowing'' limit when $\gamma_{\mu}B_\mathrm{loc}^\mathrm{rms}\tau_{c} \ll 1$, where $\gamma_{\mu} = 8.5162 \times 10^8~\text{s}^{-1}~\text{T}^{-1}$ is the gyromagnetic ratio of the muon and $B_\mathrm{loc}^\mathrm{rms} = \langle B_\mathrm{loc}^2(t)\rangle^{1/2}$ is the rms value of the fluctuating local field. This results in an exponential decay of $G(t)$:
\begin{equation}\label{Eq:simpleexp}
 G(t) = \exp(-\lambda_d t),
\end{equation}
where the dynamic relaxation rate~$\lambda_d$ is of the order of $(\gamma_{\mu}B_\mathrm{loc}^\mathrm{rms})^2\tau_{c}$, i.e., proportional to $\tau_c$.
In ZF $\lambda_d = 2(\gamma_{\mu}B_\mathrm{loc}^\mathrm{rms})^2\tau_{c}$~\cite{yaouanc2011muon}. In a LF experiment in a field~$H_L$, the field dependence of $\lambda_d$ is given by the Redfield relation~\cite{yaouanc2011muon, slichter2013principles}
\begin{equation}\label{Eq:Redfield}
 \lambda_d(H_L,\tau_c) = \frac{2(\gamma_{\mu}B_\mathrm{loc}^\mathrm{rms})^{2}\tau_{c}}{(\gamma_{\mu}H_L\tau_{c})^2+1}.
\end{equation}
This field dependence has been observed in a number of systems, including YBCO~\cite{Zhang2018Discovery} and the heavy fermion superconductor PrOs$_{4}$Sb$_{12}$~\cite{Aoki2003Time-reversal}.

In wTF-$\mu$SR in a paramagnet, the muon spins precess at close to the Larmor frequency~$\omega_\mathrm{TF}$ in the applied field, often without strong depolarization. The observed asymmetry of this signal is the sum of contributions from the sample and from those muons that miss the sample and stop in the silver sample holder~\footnote{The holder in a $\mu$SR spectrometer is usually made of pure silver, because of its good thermal conductivity and because the absence of strong nuclear magnetism simplifies its $\mu$SR signal.}. If a distribution of strong internal fields due to static magnetism (with or without magnetic order) results in rapid relaxation of the sample muon polarization, then the sample wTF signal precessing at $\omega_\mathrm{TF}$ is lost, and any signal at late times is due only to the ``silver background'' signal. This behavior is useful in two ways: (1)~the temperature dependence of the initial late-time asymmetry yields the transition temperature~\cite{yaouanc2011muon}, and (2)~as long as the muon beam remains constant, the value of the silver background asymmetry can be subtracted from the raw data to give the sample asymmetry.

\section{RESULTS} \label{sec:res}

\subsection{\boldmath AFM state} \label{sec:AFMstate}

$\mu$SR in a system with static magnetism is often complicated, due to the presence of both static and fluctuating components at muon sites and a large number of such sites in transition metal oxides~\cite{[{See, for example, }] Bernal19Charge}. Franke \emph{et~al.}~\cite{Franke2011Measurement} reported a ZF-$\mu$SR study of Sr$_2$IrO$_4$, which revealed multiple precession signals and spin reorientation in the AFM phase. The focus of the present paper is on results at temperatures above $T_N$, and we only note here that ZF-$\mu$SR spectra from the AFM phases of Rh-doped samples exhibit strongly damped oscillations at frequencies comparable to those in undoped Sr$_2$IrO$_4$.

\begin{figure} [ht]
 \includegraphics[width=0.45\textwidth]{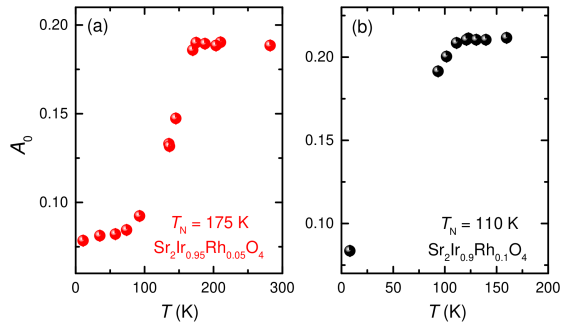}
 \caption{\label{fig:asy4TN} Temperature dependence of the late-time initial asymmetry $A_0$ at $\omega_L$ from wTF-$\mu$SR asymmetry spectra (TF = 20~mT) in (a) Sr$_2$Ir$_{0.95}$Rh$_{0.05}$O$_4$ and (b) Sr$_2$Ir$_{0.9}$Rh$_{0.1}$O$_4$ (data taken at J-PARC).}
\end{figure}

As discussed in Sec.~\ref{sec:musrtech}, wTF-$\mu$SR can be used to determine the temperature of a transition from a paramagnetic phase to one with static magnetism. Figure~\ref{fig:asy4TN} shows the temperature dependence of the initial asymmetry~$A_0$ of the signal precessing at $\omega_L$ in Sr$_2$Ir$_{1-x}$Rh$_x$O$_4$, $x = 0.05$ and 0.1, from fits to the late-time wTF spectra.
Asymmetry loss corresponding to the onset of AFM order is observed below N\'eel temperatures~$T_N =$ 175(5)~K and 110(5)~K for $x = 0.05$ and 0.1, respectively. These results agree well with previous determinations of $T_N$, as discussed in Sec.~\ref{sec:maxima}.

\subsection{\boldmath Paramagnetic state}

\subsubsection{Muon spin relaxation in ZF and low-field LF \label{sec:static}}

\begin{figure} [ht]
 \includegraphics[width=0.45\textwidth]{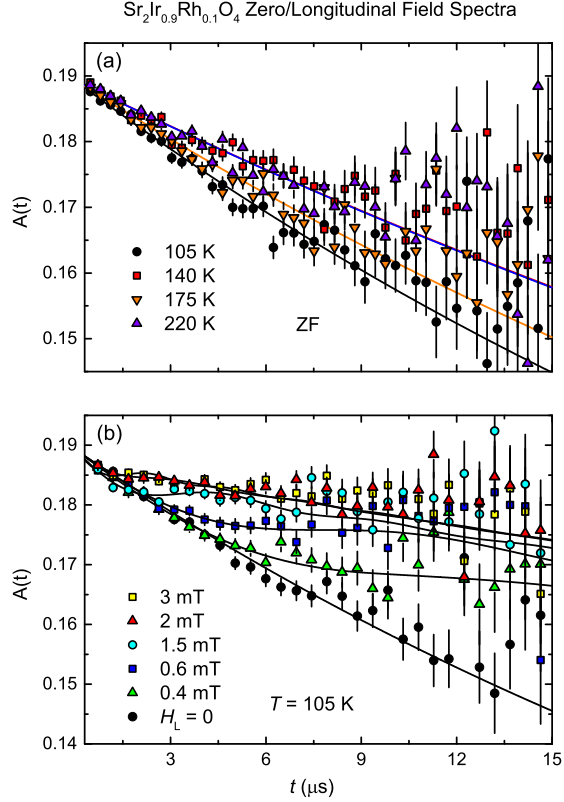}
 \caption{(a)~ZF-$\mu$SR asymmetry time spectra from Sr$_2$Ir$_{0.9}$Rh$_{0.1}$O$_4$ at various temperatures. (b)~low-field LF-$\mu$SR spectra from Sr$_2$Ir$_{0.9}$Rh$_{0.1}$O$_4$ at 105~K (data taken at ISIS). Curves: fits of Eqs.~(\ref{Eq:simpleexp}) and Eq.~(\ref{Eq:dmpdstat}) to the ZF and low-field LF data, respectively. The weakly-relaxing signal from muons that stop in the silver backing plate has not been subtracted. The curves for 140~K and 220~K data in (a) and 2~mT and 3~mT data in (b) are nearly coincident.}
 \label{fig:wLFspct}
\end{figure}

Figure~\ref{fig:wLFspct}(a) shows ZF-$\mu$SR asymmetry time spectra from Sr$_2$Ir$_{0.9}$Rh$_{0.1}$O$_4$ at four different temperatures.
All can be well fit by a simple exponential time dependence [Eq.~(\ref{Eq:simpleexp})]. In paramagnets muon spin relaxation typically varies monotonically with temperature, whereas here the muon spins depolarize faster at 175~K than at higher (220~K) and lower (140~K) temperatures.

Equations~(\ref{Eq:KTLor}) and (\ref{Eq:simpleexp}) give equally good fits to the ZF data, because they both exhibit exponential decay at the early times accessible to the experiments. To decide between the functional forms, low-field LF-$\mu$SR experiments were preformed. As shown in Fig.~\ref{fig:wLFspct}(b), the relaxation is slowed by the application of LF, the spectra exhibit strongly damped oscillations at the LF frequency, and the initial slopes of the spectra are field-independent. These are all characteristic properties of a static LF KT function~\cite{Hayano19179Zero}. There is hardly any difference between spectra at $\mu_0H_L$ = 2~mT and 3~mT, as expected from complete static-field decoupling. There is still significant relaxation at these fields, however, which must therefore be dynamic in nature.

The exponentially-damped static LF Lorentzian KT function
\begin{equation}\label{Eq:dmpdstat}
 G(t) = \exp(-\lambda_d t)G_\mathrm{LF}^\mathrm{Lor}(\lambda_s,H_L,t),
\end{equation}
models the combined effects of static (nuclear) and dynamic (electronic) local fields that coexist as vector sums at muon sites in a paramagnet. Here $G_\mathrm{LF}^\mathrm{Lor}(\lambda_s,H_L,t)$ describes static relaxation in LF for a Lorentzian local field distribution~\cite{Uemura1985Muon}. Fits of Eq.~(\ref{Eq:dmpdstat}) to the data are shown in Fig.~\ref{fig:wLFspct}(b), and describe the data well at all fields assuming a field-independent $\lambda_d$~\footnote{According to the Redfield relation [Eq.~(\ref{Eq:Redfield})], $\lambda_d$ is suppressed for fields $\gtrsim 1/\gamma_\mu \tau_c$; cf.\ Sec.~\ref{sec:Fdep}}.

The slow overall relaxation and exponential initial time dependence of $G_\mathrm{ZF}^\mathrm{Lor}(\lambda_s,t)$ render the two relaxation rates in Eq.~(\ref{Eq:dmpdstat}) highly correlated statistically in the ZF data. Furthermore, the constant term in $G_\mathrm{ZF}^\mathrm{Lor}(\lambda_s,t)$ [Eq.~(\ref{Eq:KTLor})] becomes less than 1/3 if the static fields are preferentially oriented perpendicular to the initial muon polarization; ZF data from the AFM state of Sr$_2$Ir$_{0.9}$Rh$_{0.1}$O$_4$ (not shown) indicate that this is the case. These circumstances make the observed ZF relaxation close to exponential, as shown in Fig.~\ref{fig:wLFspct}(a). The ZF exponential depolarization rate $\lambda_\mathrm{ZF}$ is contributed to by both static and dynamic fields [from Eqs.~(\ref{Eq:KTLor}) and (\ref{Eq:dmpdstat}), $\lambda_\mathrm{ZF} = \textstyle{\frac{4}{3}}\lambda_s + \lambda_d$ at early times]. We note that from Fig.~\ref{fig:wLFspct}(b) $\lambda_\mathrm{ZF} \gg\lambda_d $ from the high-field data~\footnote{$\lambda_\mathrm{LF} = \lambda_d$ at high fields because $G_\mathrm{LF}^\mathrm{Lor}(\lambda_s,H_L,t) \to 1 \text{ for } H_L \gg \lambda_s/\gamma_\mu$}, implying that $\lambda_\mathrm{ZF}$ is dominated by static relaxation.

\subsubsection{Temperature dependences of the relaxation rates \label{sec:maxima}}

\begin{figure} [ht]
 \includegraphics[width=0.48\textwidth]{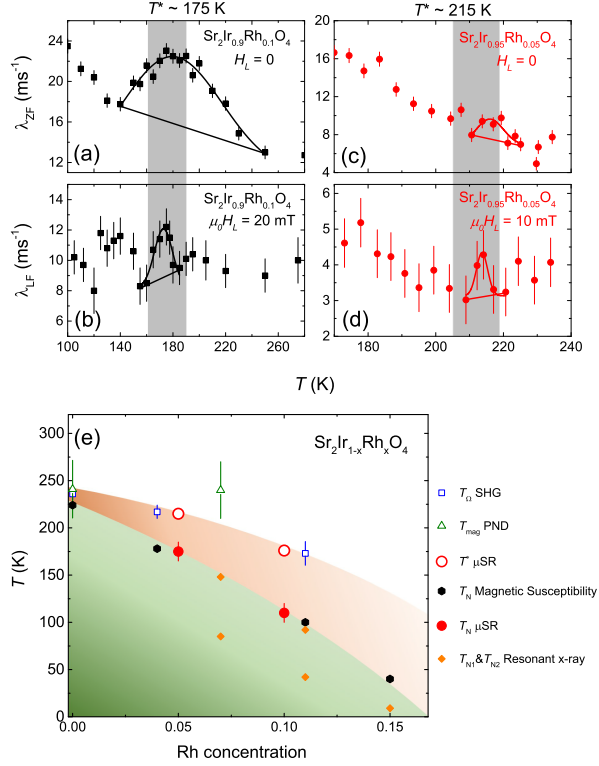}
 \caption{\label{fig:tmpphas} Temperature dependences of ZF and LF muon spin relaxation rates in Sr$_{2}$Ir$_{1-x}$Rh$_{x}$O$_{4}$. (a) $x = 0.1$, $H_L = 0$ (data taken at ISIS). (b) $x = 0.1$, $\mu_0H_L = 20$~mT\@ (data taken at PSI). (c) $x = 0.05$, $H_L = 0$ (data taken at J-PARC). (d) $x = 0.05$, $\mu_0H_L = 10$~mT (data taken at J-PARC)\@. Straight lines: baselines for statistical analysis (Sec.~\ref{sec:statistical}). Curves: guides to the eye. Gray areas: uncertainty in $T^{\ast}$ from previous reports~\cite{Zhao2015Evidence,Jeong2017Time-reversal}. (e) Magnetic phase diagram of Sr$_{2}$Ir$_{1-x}$Rh$_{x}$O$_{4}$. Filled circles: N\'eel temperatures~$T_N$ from $\mu$SR (this work). Hexagons: $T_N$ from magnetic susceptibility~\cite{Qi2012Spin}. Diamonds: long-range and short-range $T_N$ from resonant x-ray diffraction~\cite{Clancy2014Dilute}. Open circles: hidden order transition temperature~$T^{\ast}$ from $\mu$SR (this work). Squares: $T_{\Omega}$ from second harmonic generation~\cite{Zhao2015Evidence}. Triangles: $T_{mag}$ from polarized neutron diffraction~\cite{Jeong2017Time-reversal}.}
\end{figure}

The temperature dependences of $\lambda_\mathrm{ZF}$ for $x =$ 0.1 and 0.05 are shown in Figs.~\ref{fig:tmpphas} (a) and (c).
Maxima in $\lambda_\mathrm{ZF}(T)$ are observed at temperatures~$T^\ast \approx$ 175~K and 215~K for $x =$ 0.1 and 0.05, respectively. The amplitude and width of the maximum for $x = 0.1$ [Fig.~\ref{fig:tmpphas}(a)] are much larger than for $x = 0.05$ [Fig.~\ref{fig:tmpphas}(c)], however, which suggests that the maxima have different origins. As noted above, $\lambda_\mathrm{ZF}$ is due to both static and dynamic fields. We attribute the broad relaxation rate maximum in Fig.~\ref{fig:tmpphas}(a) to slow muon diffusion and trapping by substitutional disorder, as has been observed in Sc-doped SrZrO$_3$~\cite{Hempelmann1998}, another transition-metal oxide. This is discussed in detail in Sec.~\ref{sec:originofpeak}. We further argue that the narrow relaxation-rate maximum at 215~K for $x = 0.05$ [Fig.~\ref{fig:tmpphas}(c)] is most probably due to critical slowing down at the hidden-order phase transition, similar to the situation in YBCO~\cite{Zhang2018Discovery}.

Data were also taken in longitudinal fields strong enough to decouple the static component but too weak to affect the relaxation rate. The observed relaxation is simple exponential. Figures~\ref{fig:tmpphas}(b) and (d) give the temperature dependences of the relaxation rates~$\lambda_\mathrm{LF}$ in Sr$_2$Ir$_{1-x}$Rh$_{x}$O$_4$, $x = 0.1$ and 0.05, for $\mu_0H_L = 20$~mT and 10~mT, respectively. Maxima in $\lambda_\mathrm{LF}(T)$ were observed at $T^\ast \approx$ 175~K and 215~K, respectively. The statistical significance of the LF peak for $x = 0.05$ is marginal (cf.\ Sec.~\ref{sec:statistical}), but it is at the same temperature as the ZF peak.

For $x = 0.1$, the ZF rates are greater than the LF rates and the maximum is much broader. Thus the ZF peak is mainly static and is easily decoupled, whereas the peak in 20-mT LF is dynamic in origin. We note that in this interpretation it is a coincidence that the ZF and LF maxima occur at almost the same temperature; they are caused by different mechanisms and are unrelated. The situation is discussed in detail in Sec.~\ref{sec:originofpeak}.

From the proportionality between $\lambda_d$ and the fluctuation correlation time~$\tau_{c}$ (cf.\ Sec.~\ref{sec:musrtech}), the LF relaxation rate maxima suggest critical slowing down of dynamic magnetic fluctuations. Values of $T^{\ast}$ from Figs.~\ref{fig:tmpphas}(b)--(d) are plotted on the magnetic phase diagram of Sr$_{2}$Ir$_{1-x}$Rh$_{x}$O$_{4}$ in Fig.~\ref{fig:tmpphas}(e). They agree well with symmetry-breaking hidden-order transition temperatures~$T^\ast$ obtained from SHG~\cite{Zhao2015Evidence} and PND~\cite{Jeong2017Time-reversal} experiments, suggesting that the relaxation-rate maxima are due to slowing down of the hidden-order fluctuations and a hidden magnetic order transition at $T^{\ast}$. Values of $T_N$ from Fig.~\ref{fig:asy4TN} are also plotted in Fig.~\ref{fig:tmpphas}(e), and agree well with dc magnetic susceptibility~\cite{Qi2012Spin} and resonant x-ray~\cite{Clancy2014Dilute} data.

Below $T^\ast$ $\lambda_\mathrm{ZF}$ increases with decreasing temperature, whereas $\lambda_\mathrm{LF}$ is almost temperature independent. This is consistent with the muon diffusion and trapping scenario discussed in Sec.~\ref{sec:maxima} if the diffusion is slow~\cite{Luke1991Muon}. We do not understand the possible second peak at $\sim$140~K in Fig.~\ref{fig:tmpphas}(b).

\subsubsection{Field dependence of $\lambda_\mathrm{LF}$} \label{sec:Fdep}

\begin{figure} [ht]
 \includegraphics[width=0.45\textwidth]{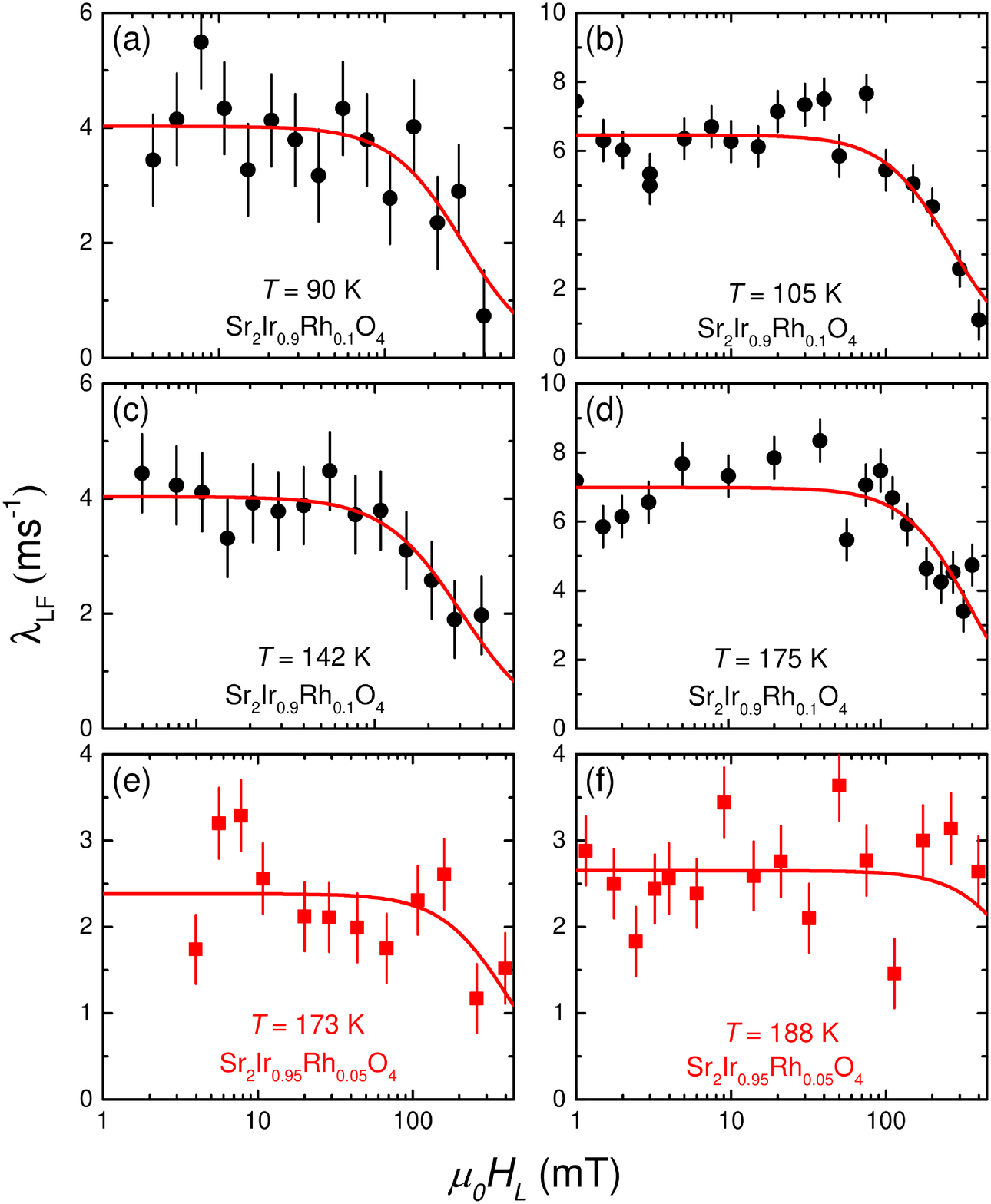}
 \caption{\label{fig:flddep} Field dependence of LF relaxation rate~$\lambda_\mathrm{LF}(H_{L})$ in Sr$_2$Ir$_{1-x}$Rh$_{x}$O$_4$, $\mu_0H_L \geqslant 2$~mT\@. (a)~$x = 0.1$, $T = 90$~K (data taken at J-PARC)\@. (b)~$x = 0.1$, $T = 105$~K (data taken at ISIS)\@. (c)~$x = 0.1$, $T = 142$~K (data taken at J-PARC)\@. (d)~$x = 0.1$, $T = 175$~K (data taken at ISIS)\@. (e)~$x = 0.05$, $T = 173$~K (data taken at J-PARC)\@. (f) $T = 188$~K (data taken at J-PARC)\@. Curves: fits of Eq.~(\ref{Eq:Redfield}) to the data [$B_\mathrm{loc}^\mathrm{rms}$ was fixed at 1.2~mT for (f)~$x = 0.05$, $T = 188$~K]. }
\end{figure}

LF-$\mu$SR experiments above 2~mT were carried out in Sr$_{2}$Ir$_{1-x}$Rh$_{x}$O$_{4}$, $x = 0.05$ and $0.1$, at a number of different temperatures. As shown in Figs.~\ref{fig:flddep}(a)--(f), the field dependences of $\lambda_\mathrm{LF}$ are well described by the Redfield relation [Eq.~(\ref{Eq:Redfield})]~\footnote{We note that in Fig.~\ref{fig:flddep}(a), $T = 90$~K is lower than the nominal value~$T_N = 110$~K for $x = 0.1$. However, Fig.~\ref{fig:asy4TN}(b) shows only a small loss of late-time asymmetry at this temperature. Thus 110~K should be considered as the onset of a spatial distribution of N\'eel temperatures, and the data of Fig.~\ref{fig:flddep}(a) characterize paramagnetic regions.} .
We note that values of $\lambda_\mathrm{LF}$ differ between Figs.~\ref{fig:tmpphas} and \ref{fig:flddep} (error bars in these figures are statistical only). This systematic error is due to differences in small offsets between facilities and between experimental runs at the same facility, particularly in the characterization of the silver background asymmetry noted in Sec.~\ref{sec:AFMstate}. It has been noted previously~\cite{Bueno2011Longitudinal} that such effects can cause overall uncertainties of a few ms$^{-1}$ in rates as small as those in these experiments. The offsets are unlikely to change during a given field or temperature scan, so that the reported temperature and field dependences are qualitatively significant.

The resulting fit parameters are given in Table~\ref{Tab:Redfieldparams}\@.
\begin{table}
\caption{rms muon local fields $B_\mathrm{loc}^\mathrm{rms}$ and correlation times $\tau_\mathrm{c}$ for Sr$_2$Ir$_{0.9}$Rh$_{0.1}$O$_4$ and Sr$_2$Ir$_{0.95}$Rh$_{0.05}$O$_4$ from fits of Eq.~(\ref{Eq:Redfield}) to the data of Fig.~\ref{fig:flddep}.} \label{Tab:Redfieldparams}
 \begin{ruledtabular}
 \begin{tabular}{cccc}
 Rh concentration & $T$ (K) & $B_\mathrm{loc}^\mathrm{rms}$ (mT) & $\tau_\mathrm{c}$ (ns) \\
 \hline
 0.1 & 90~K & 0.84(8) & 4.0(9) \\
 0.1 & 105~K & 1.0(7) & 4(1) \\
 0.1 & 120~K & 0.9(1) & 2.8(8) \\
 0.1 & 142~K & 0.85(4) & 3.8(4) \\
 0.1 & 160~K & 1.3(3) & 1.5(7) \\
 0.1 & 175~K & 1.3(1) & 3.0(4) \\
 0.05 & 173~K & 0.8(1) & 3(1) \\
 0.05 & 188~K & 1.2(fix) & 1(1)
 \end{tabular}
 \end{ruledtabular}
\end{table}
For $x = 0.1$ $B_\mathrm{loc}^\mathrm{rms}$ is around 1~mT below 175~K, and $\tau_c$ is in the range~1.5--4~ns. The relaxation is in the extreme motional narrowing limit: $\gamma_{\mu}B_\mathrm{loc}^\mathrm{rms}\tau_{c} \sim 0.004$, confirming that the muon local field is dynamic in nature. At 227~K $\lambda_\mathrm{LF}$ is field independent up to 400~mT (data not shown), suggesting a shorter correlation time at this temperature, although its value for $\mu_0H_L$ = 20~mT is comparable to those at lower temperatures [Fig.~\ref{fig:tmpphas}(b)].

For $x = 0.05$ the field dependences of $\lambda_\mathrm{LF}$ are not as clear as for $x = 0.1$, due to the smaller values of the rates and the likelihood that $\tau_c$ is too short to allow the decrease in $\lambda_\mathrm{LF}$ for $\gamma_\mu H_L\tau_c > 1$ [Eq.~(\ref{Eq:Redfield})] to be well resolved with the highest available field. The 188-K data are hard to fit with both $B_\mathrm{loc}^\mathrm{rms}$ and $\tau_{c}$ free without data from higher fields ($\mu_0H_L > 400$~mT); $B_\mathrm{loc}^\mathrm{rms}$ was fixed at 1.2~mT to obtain an estimated correlation time that is shorter than 2~ns.

The observed small relaxation rates are close to the limit of the $\mu$SR technique~\cite{Bueno2011Longitudinal}, and the uncertainties of the fitted parameters from Eq.~(\ref{Eq:Redfield}) are large. The data of Table~\ref{Tab:Redfieldparams} do not exhibit a systematic temperature dependence, and thus we do not have direct evidence for critical slowing down from the temperature dependence of $\tau_c$. However, rough values of $B_\mathrm{loc}^\mathrm{rms}$ and $\tau_{c}$ are obtained for both Rh concentrations.

\section{DISCUSSION}

\subsection{\boldmath Comparison with other experiments and theory}

As noted above, the observed values of the peak temperatures~$T^\ast$ are in good agreement with the results of SHG~\cite{Zhao2015Evidence} and PND~\cite{Jeong2017Time-reversal} experiments. This and the magnitude of the observed relaxation resemble the results of Zhang \emph{et~al.} in YBCO~\cite{Zhang2018Discovery}, suggesting a similar origin in both systems.

Values of $B_\mathrm{loc}^\mathrm{rms}$ from Table~\ref{Tab:Redfieldparams} can be compared with PND results. Fauqu\'{e} \emph{et~al.}~\cite{Fauque2006Magnetic} reported that in YBCO the neutron scattering cross section from IUC order is $\sim$1--2 mbarn/formula unit, from which they deduce an ordered magnetic moment of 0.05--0.1 $\mu_B$. Based on dipolar lattice-sum calculations~\cite{MacDougall2008Absence, Zhang2018Discovery}, $B_\mathrm{loc}^\mathrm{rms} \approx 1$--1.5~mT for 0.1-$\mu_B$ IUC moments. For Sr$_2$Ir$_\mathrm{1-x}$Rh$_x$O$_4$, the magnetic cross-section of hidden magnetic order was estimated as $\sim$ 2~mbarn/formula unit~\cite{Jeong2017Time-reversal}. The 1-mT values of $B_\mathrm{loc}^\mathrm{rms}$ in these materials (Table~\ref{Tab:Redfieldparams}) are therefore roughly consistent with the PND results~\cite{Jeong2017Time-reversal}.

The magnitude of $B_\mathrm{loc}^\mathrm{rms}$ for hidden magnetic order is $\sim$ 20 times smaller than the static local field at muon sites below $T_{N}$~\cite{Franke2011Measurement}. This might be expected; in general rms fluctuating fields are smaller than static fields in magnetically-ordered states. Here, however, there is no evidence for static magnetization for $T_N < T < T^{\ast}$ (see below), and the discrepancy is an indication that the sources of the two fields are quite different. A difference (individual local moments vs loop currents) is expected within the loop-current theory~\cite{Varma14Pseudogap} discussed below. The PND study reported magnetic scattering for hidden order $\sim$5 times smaller than for AFM order~\cite{Jeong2017Time-reversal}, which accounts for part of the difference. Further comparison would require improved data and better knowledge of expected loop-current/$\mu^+$ coupling.

Results from the $\mu$SR experiments differ from those from SHG~\cite{Zhao2015Evidence} and PND~\cite{Jeong2017Time-reversal} in two important ways. With respect to determining the hidden-order transition temperature, SHG measures a parameter that characterizes the level of rotational symmetry breaking, and PND observes the magnitude of a spontaneous magnetization. Both signals increase with decreasing temperature below the hidden-order transition. In contrast, as discussed in detail below in Sec.~\ref{sec:BPPpeaks}, the peak in the $\mu^+$ relaxation rate at $T^\ast$ reflects a maximum in the correlation time~$\tau_c$. The fluctuations are quasistatic on the SHG and PND time scales (${\sim}10 ^{-11}$~s), which are much order are shorter than $\tau_c \approx 10^{-9}$~s from $\mu$SR (Table~\ref{Tab:Redfieldparams}).

With respect to distinguishing AFM order from hidden order, SHG experiments reveal a different rotational symmetry, and PND patterns rule out AFM order for $T_N < T < T^{\ast}$. In ZF and weak-LF $\mu$SR experiments, static magnetism (ordered or disordered) results in oscillations or rapid damping, depending on the width of local field distribution. $\mu$SR experiments do indeed observe static AFM order below $T_N$~\cite{Franke2011Measurement}, but for $T_N < T < T^{\ast}$ no static magnetism of any kind was found in this work. Similarly, as noted in the Introduction, no static magnetism was observed in the pseudogap phase of YBCO~\cite{Zhang2018Discovery}.

In the loop-current theory of Varma~\cite{Varma14Pseudogap}, loop-current order forms in domains with a finite correlation length and consequent low-frequency excitations at all temperatures. These result in fluctuations of the IUC magnetic order that are quasistatic on the PND and SHG time scales, but prevent truly static order. This prediction is consistent with our $\mu$SR results, as noted above. Quantitative comparison with the loop-current theory must await its further development.

To within considerable errors, the correlation time~$\tau_c$ in Sr$_2$Ir$_\mathrm{1-x}$Rh$_x$O$_4$ exhibits a slight increase with $x$ (Table~\ref{Tab:Redfieldparams}). This disagrees with the loop-current theory~\cite{Varma14Pseudogap}, in which the fluctuation frequency increases with the decreased correlation length~$\xi$ due to increased doping. The lack of strong doping dependence may be associated with competition between the decrease of $\xi$ and the observed decrease of $T^\ast$ with increasing Rh concentration [Fig.~\ref{fig:tmpphas}(e)], if the latter signals a decrease of the fluctuation energy scale.

\subsection{\boldmath Origin of the relaxation rate maxima} \label{sec:originofpeak}

\subsubsection{Static relaxation in ZF}

\paragraph{Nuclear dipolar fields.} At all temperatures the ZF relaxation rate is dominated by a static field distribution, due putatively to nuclear magnetic moments. To confirm this, lattice-sum calculations of rms nuclear dipolar fields have been carried out for two candidate muon sites in the Sr$_2$IrO$_4$ crystal structure: (a)~halfway between nearest-neighbor O(1) sites in the Ir/Rh-O plane, and (b)~halfway between nearest-neighbor O(1)-O(2) sites~\cite{huang1994neutron}. Although muon stopping sites in Sr$_2$IrO$_4$ have not been determined experimentally or calculated \emph{ab initio}~\cite{[{See, for example, }] Bonfa2016toward}, in oxides they are typically close to O$^{2-}$ ions; fields at these sites can be taken as representative.

The calculations yield ZF nuclear dipolar relaxation rates (a) 26.2~ms$^{-1}$ and (b) 38.6~ms$^{-1}$ for initial muon polarization parallel to the $c$ axis of the crystal. The observed value of $\lambda_\mathrm{ZF}$ for Sr$_2$Ir$_{0.9}$Rh$_{0.1}$O$_4$ is about 20~ms$ ^{-1}$ at 105~K [Fig.~\ref{fig:tmpphas}(a)], of the same order as the calculations. This is evidence that the static relaxation in ZF is due to nuclear dipolar fields.

The calculated relaxation assumes a Gaussian field distribution, however, whereas the low-field behavior of the muon relaxation is well described by a Lorentzian distribution (Sec.~\ref{sec:static}). Thus the comparison is only qualitative. A Lorentzian nuclear dipolar field distribution was also observed in the cuprate-analog nickelate~La$_4$Ni$_3$O$_8$~\cite{Bernal19Charge}, and attributed to a large number of candidate muon stopping sites in the unit cell. This is also likely to be the situation in Sr$_2$Ir$_{0.9}$Rh$_{0.1}$O$_4$, given the complex crystal structures of these materials.

\paragraph{Muon diffusion and trapping.} These are in general complex phenomena~\cite{schenck1985muon, brewer1994encyclopedia, yaouanc2011muon, Karlsson1995Quantum}. In a paramagnet without quantum diffusion, a rough description is as follows: at low temperatures, positive muons are stationary at interstitial sites, and the muon relaxation by nuclear dipolar fields is static. With increasing temperature, muons begin to ``hop'' from site to site with increasing frequency by an activated process. The dipolar fields experienced by muons now fluctuate at the hop rate, and become a dynamic relaxation mechanism. This leads to a decrease of rate with increasing temperature due to motional narrowing. As shown in Figs.~\ref{fig:tmpphas}(a) and (c), in Sr$_2$Ir$_{1-x}$Rh$_x$O$_4$ $\lambda_\mathrm{ZF}$ exhibits such a decrease underneath the relaxation rate maxima for both concentrations.

Crystal defects can trap diffusing muons by increasing the binding energy at defect sites. At intermediate temperatures muon diffusion is slow, and most muons decay before reaching such a trap. With increasing temperature, muons are more likely to arrive at a trap. Then, if the muon spin relaxation is fast enough at the trap site, the observed ZF relaxation rate~$\lambda_\mathrm{ZF}(T)$ passes through a minimum with increasing temperature. At still higher temperatures muons escape from traps, so that $\lambda_\mathrm{ZF}(T)$ passes through a maximum and subsequently decreases.

This is qualitatively what is observed in Sr$_2$Ir$_{0.9}$Rh$_{0.1}$O$_4$ [Fig.~\ref{fig:tmpphas}(a)]. In Sr$_2$Ir$_{0.95}$Rh$_{0.05}$O$_4$ $\lambda_\mathrm{ZF}(T)$ is similar to that for $x = 0.1$ at the same temperatures, except for the peak at $T^\ast = \sim$215~K\@. This would be expected if the trapping maximum were concentration-independent, with the consequence that for $x = 0.05$ the peak at $T^\ast$ in ZF is not concealed under the trapping maximum (which may be partially observed at 170--180~K, but we note that $T_N$ is close to this temperature range). LF decoupling is present at all temperatures; relaxation rates in 20 and 10 mT are considerably smaller than in ZF (Fig.~\ref{fig:tmpphas}). The situation may be similar to that in copper for low hopping rates~\cite{Luke1991Muon}.

Hempelmann \emph{et~al.}~\cite{Hempelmann1998} reported similar results for the transition-metal oxide Sc-doped SrZrO$_3$, where a broad maximum in the muon relaxation rate at $\sim$220~K was attributed to muon diffusion and trapping at Sc ions with subsequent relaxation by $^{45}$Sc nuclear dipolar fields. Rapid decoupling in LF indicated that these fields are quasistatic. The $^{45}$Sc nuclear isotope is 100\% abundant, with a large nuclear moment (4.756$\mu_N$) and higher isotopic abundance than other nuclei in Sc-doped SrZrO$_3$, and the observation that $\lambda_\mathrm{ZF}(T)$ at the maximum is considerably faster than at low temperatures is evidence for ascribing the peak to muon trapping. The temperature of the maximum was found to be independent of Sc doping level.

$^{103}$Rh nuclear moments in Sr$_2$Ir$_{1-x}$Rh$_x$O$_4$ are considerably smaller than $^{45}$Sc moments in Sc-doped SrZrO$_3$, as are the observed values of $\lambda_\mathrm{ZF}(T)$~\cite{Hempelmann1998}. The effect of Rh-ion muon trapping at temperatures near the rate maximum may simply be to localize the muon long enough to restore static relaxation; indeed, for $x = 0.1$ $\lambda_\mathrm{ZF}(T^\ast)$ is only slightly smaller than the low-temperature value [Fig.~\ref{fig:tmpphas}(a)]. No extra relaxation mechanism specific to the Rh ion is required.

We conclude that muon diffusion and trapping account for ZF relaxation in Sr$_2$Ir$_{1-x}$Rh$_x$O$_4$ (apart from the 215-K ZF peak for $x = 0.05$). With respect to comparison of ZF and LF data for $x = 0.1$, in ZF the LF peak [Fig.~\ref{fig:tmpphas}(b)] is presumably overwhelmed by the trapping peak [Fig.~\ref{fig:tmpphas}(a)]. This is not surprising since the fluctuating fields add in quadrature, thus enhancing the suppression of the smaller contribution.

\subsubsection{BPP peaks?} \label{sec:BPPpeaks}

As an alternative to the critical-slowing hypothesis, the LF maxima in Figs.~\ref{fig:tmpphas}(b) and (d) might be Bloembergen-Purcell-Pound (BPP) peaks~\cite{Bloembergen48relaxation}, i.e., maxima in Eq.~(\ref{Eq:Redfield}) as a function of $\tau_c$ at $\tau_c = 1/\gamma_\mu H_L$. In this scenario $\tau_c$ varies monotonically with temperature, and the relaxation-rate peaks at $T^\ast$ are not due to critical slowing down.

\begin{figure} [ht]
 \includegraphics[width=0.45\textwidth]{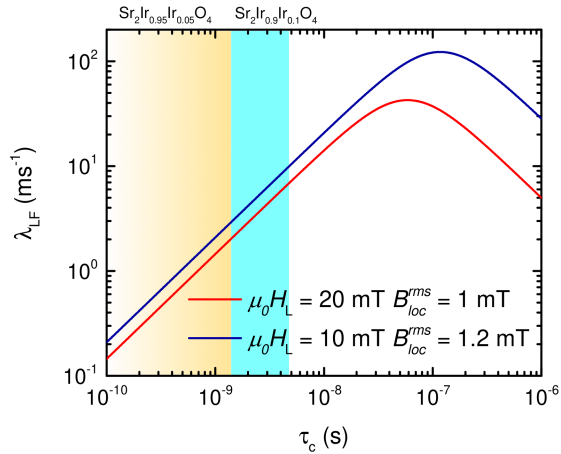}
 \caption{Calculated relaxation rate from Eq.~(\ref{Eq:Redfield}) for $\mu_0H_L$ = 20~mT, $B_\mathrm{loc}^\mathrm{rms}$ = 1~mT ($x = 0.1$, red) and $\mu_0H_L$ = 10~mT, $B_\mathrm{loc}^\mathrm{rms}$ = 1.2~mT ($x = 0.05$, blue). The ranges of estimated $\tau_c$ values are given by the light cyan area for $x = 0.1$ and the yellow area for $x = 0.05$.}
 \label{fig:tauc}
\end{figure}

We test such a hypothesis by comparing Eq.~(\ref{Eq:Redfield}) with the observed $\tau_c$. Figure~\ref{fig:tauc} shows the calculated dependence of $\lambda_\mathrm{LF}$ on $\tau_c$ from Eq.~(\ref{Eq:Redfield}) for $\mu_0H_L$ = 20~mT, $B_\mathrm{loc}^\mathrm{rms}$ = 1~mT; and $\mu_0H_L$ = 10~mT, $B_\mathrm{loc}^\mathrm{rms}$ = 1.2~mT [Figs.~\ref{fig:tmpphas}(a)--(d)].
The yellow and cyan areas represent rough estimates of the ranges of $\tau_\mathrm{c}$ values for $x = 0.05$ and 0.1, respectively. In these ranges $\lambda_\mathrm{LF}$ varies monotonically with $\tau_\mathrm{c}$. The possibility that the maxima at $T^{\ast}$ are due to a passage through the BPP peak is therefore inconsistent with the measured $\tau_c$ values.

\subsection{Statistical analysis \label{sec:statistical}}

The statistical significance of the peaks in $\lambda(T)$ [Figs.~\ref{fig:tmpphas}(b)--(d)] is not large and requires further discussion. Following the treatment of Ref.~\onlinecite{Zhang2018Discovery}, we define the inverse relative standard deviation (IRSD) (the $N$ in ``$N\sigma$'') for each peak as follows: a baseline is drawn between points chosen above and below the peak, as shown in Fig.~\ref{fig:tmpphas}. The IRSD for the peak is then $\left[ \sum_i A_i|A_i|/\sigma_i^2 \right]^{1/2}$, where $A_i$ is the point amplitude relative to the baseline, $\sigma_i$ is the point standard deviation, and the sum is over the points in the peak. The sign of $A_i$ is included to account for negative contributions.

The results are shown in Table~\ref{table:II}.
\begin{table}[ht]
\caption{IRSDs of relaxation rate maxima near $T^{\ast}$ from muon spin relaxation rate in Sr$_2$Ir$_{1-x}$Rh$_x$O$_4$. } \label{table:II}
 \begin{ruledtabular}
 \begin{tabular}{ccccc}
 x & $\mu_0H_T$ (mT) & No. of points & IRSD \\
 \hline
 0.1 & 0 & 17 & 28.2 & Fig.~\ref{fig:tmpphas}(a)\\
 & 20 & 8 & 4.4 & Fig.~\ref{fig:tmpphas}(b) \\
 \hline
 0.05& 0 & 7 & 4.8 & Fig.~\ref{fig:tmpphas}(c)\\
 & 10 & 5 & 2.2 & Fig.~\ref{fig:tmpphas}(d)\\
 \end{tabular}
 \end{ruledtabular}
\end{table}
The ZF IRSD for $x = 0.1$ is very large because the peak is dominated by muon trapping. The IRSD values for $x = 0.1$ in 20~mT and $x = 0.05$ in ZF are close to the usual criterion of 5, given the crudeness of the procedure, and are good evidence for the reality of these peaks. The IRSD value of 2.2 for $x = 0.05$ in 10~mT is not close to 5, however, and the data must be considered as suggestive but not conclusive. A significant improvement in statistics would require an impractically long time.

\section{CONCLUSIONS}

A $\mu$SR study of Sr$_2$Ir$_{1-x}$Rh$_{x}$O$_{4}$, $x = 0.05$ and 0.1, has revealed slow electronic magnetic fluctuations, which appear to be associated with the hidden magnetic order discovered in SHG~\cite{Zhao2015Evidence} and PND~\cite{Jeong2017Time-reversal} experiments. Maxima in muon relaxation rates near previously-reported hidden-order transition temperatures~$T^\ast$~\cite{Zhao2015Evidence, Jeong2017Time-reversal} are observed in ZF for $x = 0.05$ and in LF for both concentrations. These results suggest critical slowing down of magnetic fluctuations and an unconventional ordered state in which fluctuations persist for temperatures between $T^\ast$ and an AFM transition at $T_N$. The field dependences of $\lambda_\mathrm{LF}$ are described by the Redfield relation [Eq.~(\ref{Eq:Redfield})], fits to which yield estimated correlation times in the range of 1.5--4~ns for Sr$_2$Ir$_{0.9}$Rh$_{0.1}$O$_4$ and shorter than 2~ns for Sr$_2$Ir$_{0.95}$Rh$_{0.05}$O$_4$. These results are quite similar to those observed in the YBCO system~\cite{Zhang2018Discovery} and attributed to IUC order. For $x = 0.1$ the ZF muon relaxation is dominated by nuclear dipolar fields, with a broad rate maximum in the temperature dependence of the rate that is ascribed to muon diffusion and trapping.

\begin{acknowledgments}
We are grateful to C.~M. Varma and R. Kadono for helpful discussions. This research was supported by the National Key Research and Development Program of China, No.~2017YFA0303104 and No.~2016YFA0300503, the National Natural Science Foundation of China, No.~11774061, the U.S. National Science Foundation, No.~DMR/PREM-1523588, No.~HRD-1547723 and No.~DMR-1905636, and by the University of California, Riverside, Academic Senate.
\end{acknowledgments}

%\bibliographystyle{prsty}
%\bibliography{SrIrO_v8}


\end{document}